\documentclass[3p,times,procedia]{elsarticle}
\usepackage{nupha_ecrc}

\volume{00}

\firstpage{1}

\journalname{Nuclear Physics A}

\runauth{H.~Paukkunen}

\jid{nupha}

\jnltitlelogo{Nuclear Physics A}

\usepackage{amssymb}

\usepackage[figuresright]{rotating}

\usepackage[utf8]{inputenc}
\usepackage{color}
\usepackage{wrapfig}
\usepackage{mathtools}

\begin{document}

\begin{frontmatter}

%% Instructions from Editor: Please use the following \dochead only in the preprint version (e-print arXiv etc.); 
%% use empty \dochead{} when submitting to Nuclear Physics A!

\dochead{XXVIth International Conference on Ultrarelativistic Nucleus-Nucleus Collisions\\ (Quark Matter 2017)}
%\dochead{}

\title{Status of nuclear PDFs after the first LHC p--Pb run}

\author{Hannu Paukkunen}

\address{University of Jyvaskyla, Department of Physics, P.O. Box 35, FI-40014 University of Jyvaskyla, Finland \linebreak
Helsinki Institute of Physics, P.O. Box 64, FI-00014 University of Helsinki, Finland \linebreak
Instituto Galego de F\'\i sica de Altas Enerx\'\i as (IGFAE), Universidade de Santiago de Compostela, E-15782 Galicia, Spain
}

\ead{hannu.paukkunen@jyu.fi}

\begin{abstract}
In this talk, I overview the recent progress on the global analysis of nuclear parton distribution functions (nuclear PDFs). After first introducing the contemporary fits, the analysis procedures are quickly recalled and the ambiguities in the use of experimental data outlined. Various nuclear-PDF parametrizations are compared and the main differences explained. The effects of nuclear PDFs in the LHC p--Pb hard-process observables are discussed and some future prospects sketched.
\end{abstract}


\end{frontmatter}

\section{Introduction}
\label{Introduction}

\vspace{-0.1cm}
The global analysis of nuclear PDFs is theoretically founded in collinear factorization \cite{Collins:1989gx}, where the cross sections are convolutions
\begin{figure}[htb!]
\centering
\includegraphics[width=0.470\linewidth]{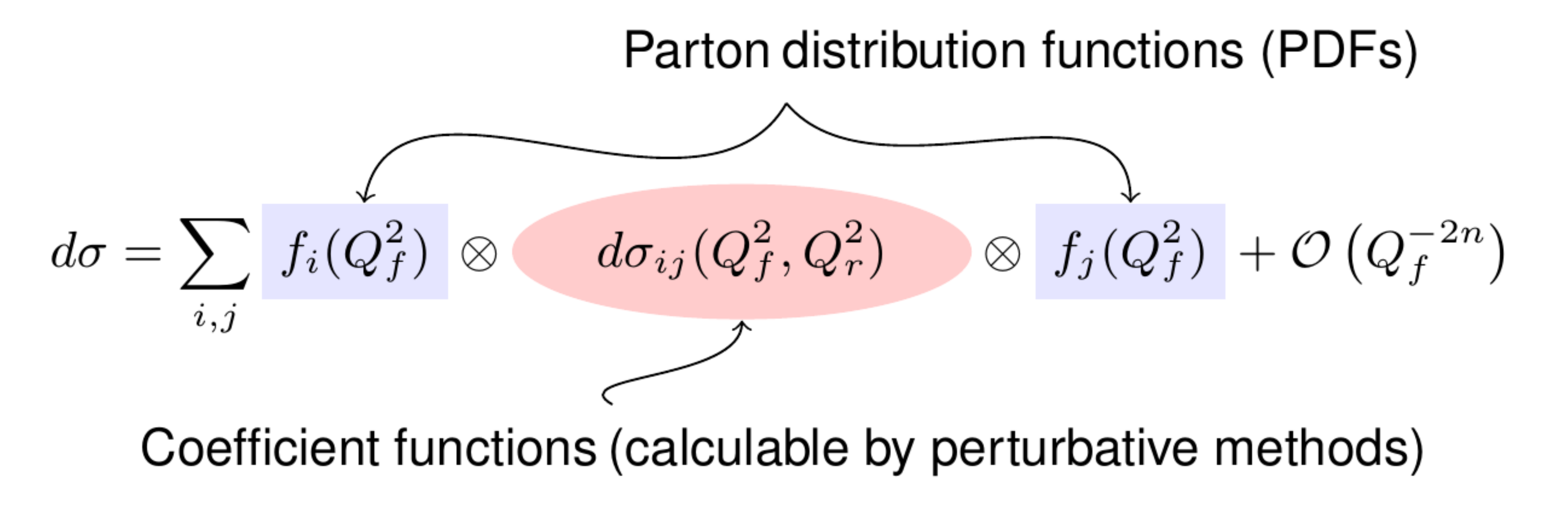}
\includegraphics[width=0.400\linewidth]{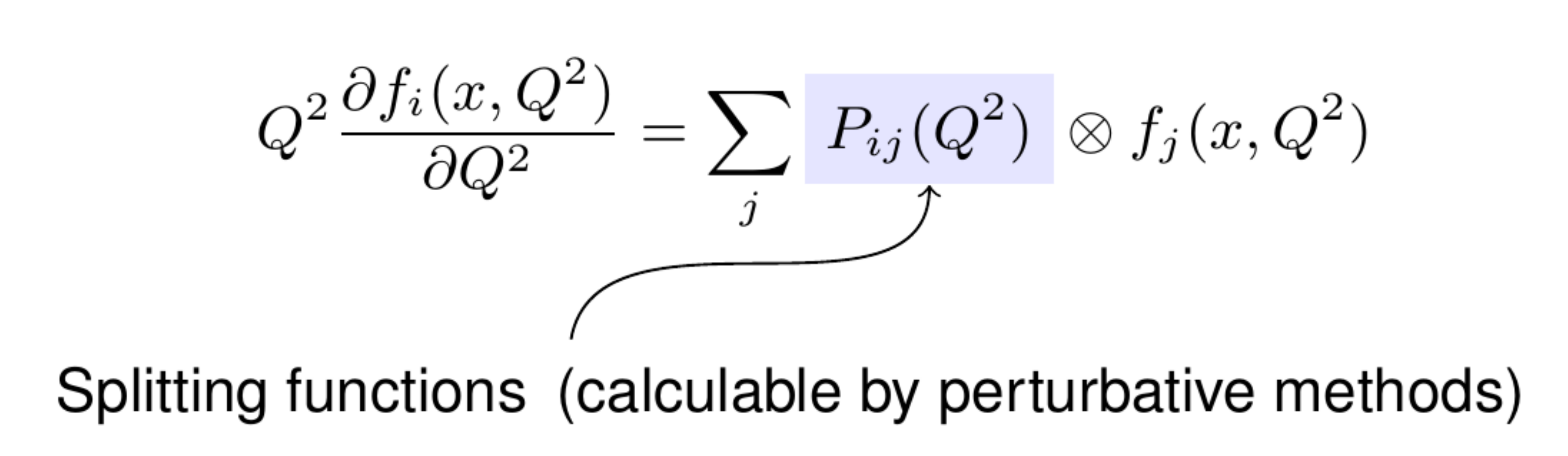}
\label{fig:fact}
\caption{Ingredients of the factorization theorem in hadronic collisions (left) and the partonic evolution equations (right).}
\end{figure} 
between non-perturbative parton distribution functions $f_i(x,Q^2_f)$ and perturbative matrix elements $d\sigma_{ij}$. While the momentum-fraction ($x$) dependence of the PDFs cannot be yet directly calculated from first principles, their scale ($Q_f^2$) dependence is given by the partonic DGLAP evolution equations where the splitting functions $P_{ij}$ can be computed by perturbative methods of QCD and electroweak theory. Fig.~\ref{fig:fact} below illustrates these ingredients. The factorization is known to break down at low $Q_f^2$ where the multi-parton interactions give rise to $Q_f^{-2n}$ power corrections \cite{Albacete:2014fwa}, particularly for large nuclei at small $x$. Factorization is also often supplemented with external models for hadronization (like the one in PYTHIA \cite{Sjostrand:2014zea}), and in extreme case even with fluid dynamical descriptions \cite{Niemi:2015qia}.

Several sets of nuclear PDFs are available --- the latest ones are listed in Table~\ref{tab:table} along with some technical details. Most of these, EPS09 \cite{Eskola:2009uj}, DSSZ \cite{deFlorian:2011fp}, nCTEQ15 \cite{Kovarik:2015cma}, EPPS16 \cite{Eskola:2016oht}, are implemented at next-to-leading order (NLO) accuracy in perturbative QCD, but also the first try towards next-to-NLO level has recently emerged \cite{Khanpour:2016pph}. All the analyses use the old fixed-target deep inelastic scattering (DIS) and proton-nucleus Drell-Yan (DY) dilepton-production data as a constraint. Most use inclusive pion production from RHIC as well. The DSSZ and EPPS16 fits are the only ones to use neutrino DIS data, but EPPS16 goes far beyond including also fixed-target DY data in pion-nucleus collisions and new LHC p--Pb data on dijet and heavy gauge-boson production. Theoretical details vary, but I should point out the movement towards a consistent inclusion of heavy-quark coefficient functions and flavor-dependent nuclear effects.

\begin{wrapfigure}{r}{0.400\textwidth}
\vspace{-0.6cm}
\includegraphics[width=1.00\linewidth]{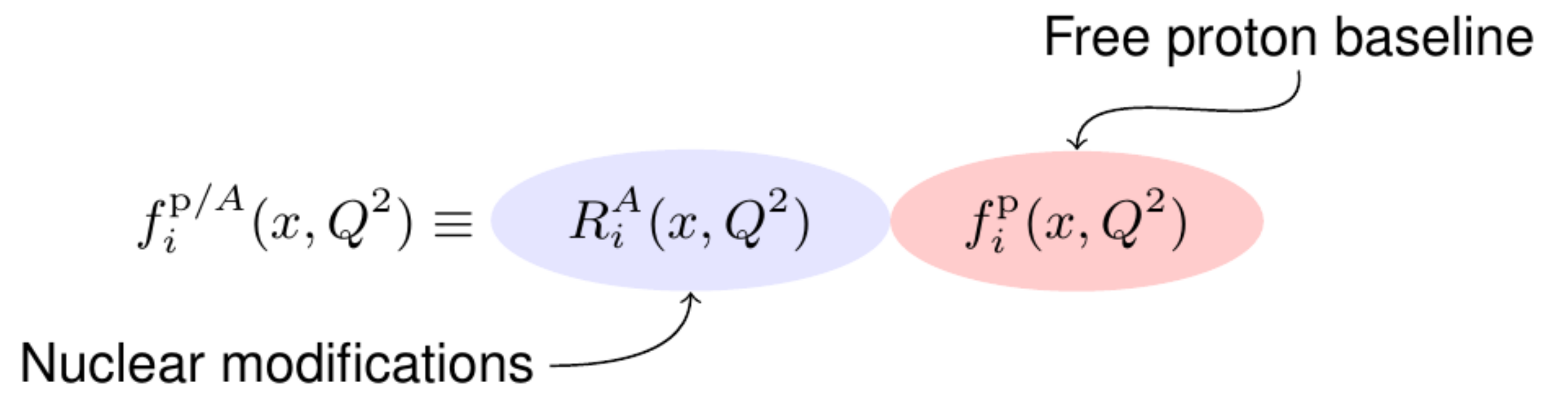} \\
\includegraphics[width=1.00\linewidth]{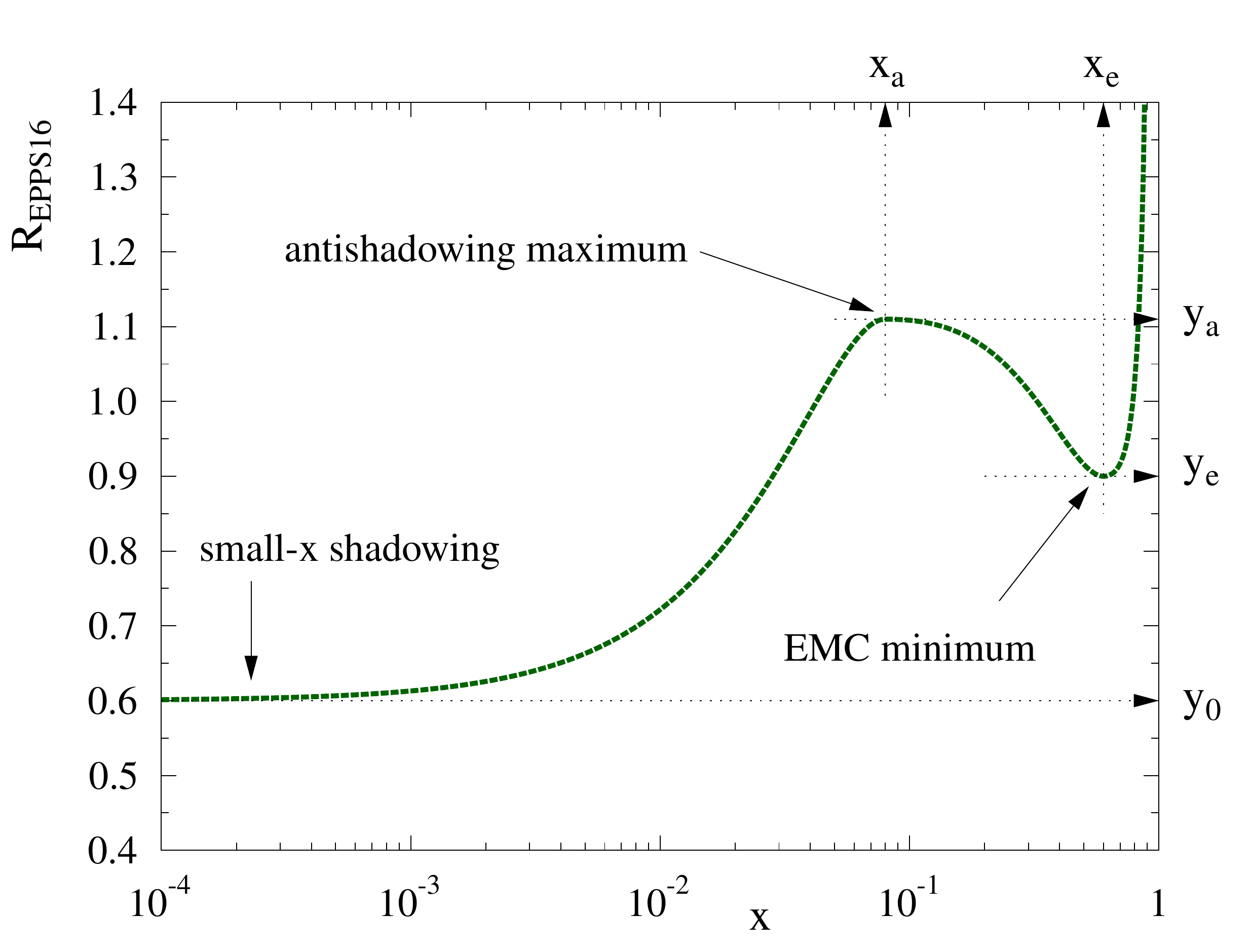}
\caption{The definition of nuclear PDFs and a typical fit function illustrated. Figure from Ref.~\cite{Eskola:2016oht}.}
\label{fig:R}
\end{wrapfigure}

{
\begin{table}
\caption{Key characteristics of the latest fits of nuclear PDFs in chronological order.}
\label{tab:table}
\begin{center}
\begin{tabular}{|c||c|c|c|c|c|}
\hline
& \textsc{eps09}  & \textsc{dssz12}	 & \textsc{ka15}	 	& \textsc{ncteq15} & \textsc{epps16} \\
\hline
\small{Order in $\alpha_s$}                   & \small{LO} \& \small{NLO} & \small{NLO}           & \small{NNLO}               & \small{NLO} & \small{NLO}        \\
\small{Neutral current DIS $\ell$+A/$\ell$+d} & \checkmark                & \checkmark                  & \checkmark                & \checkmark     &  \checkmark     \\
\small{Drell-Yan dilepton p+A/p+d}            & \checkmark                & \checkmark                  & \checkmark                & \checkmark     &  \checkmark     \\
\small{RHIC pions d+Au/p+p}                   & \checkmark                & \checkmark                  &                 &  \checkmark  &  \checkmark                 \\
\small{Neutrino-nucleus DIS}                  &                           & \checkmark                  &                 &         & \checkmark            \\
\small{Drell-Yan dilepton $\pi$+$A$}          &     &   &                           &  & { \checkmark  }                     \\                                            
\small{LHC p+Pb jet data   }                  &     &   &                           &     &  { \checkmark  }              \\
\small{LHC p+Pb W, Z data  }                  &     &  &                          &      &  { \checkmark  }               \\
&                           &                             &                           &  & \\
\small{$Q$ cut in DIS}                      & $1.3 \, {\rm GeV}$          & $1 \, {\rm GeV}$          & $1 \, {\rm GeV}$          & $2 \, {\rm GeV}$   & $1.3 \, {\rm GeV}$  \\
\small{datapoints}                            & 929                      & 1579                         & 1479                      & 708    & 1811             \\
\small{free parameters}                       & 15                        & 25                          & 16                        & 17     & 20             \\
\small{error analysis}                        & Hessian                   & Hessian                     & Hessian                   & Hessian  & Hessian        \\
\small{error tolerance $\Delta \chi^2$}       & 50                      & 30                          & not given                        & 35        & 52           \\
\small{Free proton baseline PDFs}             & \small{\textsc{cteq6.1}}   & \small{\textsc{mstw2008}}    & \small{\textsc{jr09}} & \small{\textsc{cteq6m}-like} & \small{\textsc{ct14NLO}} \\
\small{Heavy-quark effects}                    &                            & \checkmark                   &                        & \checkmark & \checkmark            \\
\small{Flavor separation}                    &                       &                         &                       & some & \checkmark \\
\small{Reference}                                 & \textcolor{blue}{{\tiny \textsc{[JHEP 0904 065]}}}                       & \textcolor{blue}{{\tiny \textsc{[PR D85 074028]}}}                         & \textcolor{blue}{{\tiny \textsc{[PR D93, 014026]}}}                       & \textcolor{blue}{{\tiny \textsc{[PR D93 085037]}}} & \textcolor{blue}{{\tiny \textsc{[EPJ C77 163]}}}  \\
\hline
\end{tabular}
\end{center}
\end{table}
}

\vspace{-0.3cm}
\section{Analysis procedures}
\label{Analysisprocedures}

\vspace{-0.2cm}
As in Fig.~\ref{fig:R}, one can always write the bound proton nuclear PDFs $f_i^{p/A}(x,Q^2)$ in terms of nuclear modifications $R_i^{A}(x,Q^2)$ and proton PDFs $f_i^{p}(x,Q^2)$. Indeed, this is what one implicitly always does --- even when $f_i^{p/A}(x,Q^2)$s are parametrized directly \cite{Kovarik:2015cma}. The point is that much of the heavier-nucleus data included in the present analyses are ratios to proton or deuteron measurements, so that the free-proton PDFs must always be supplied and the resulting nuclear PDFs are always tied to the chosen free-proton PDFs. A typical behavior of the nuclear modification $R_i^{A}(x,Q^2)$ as a function of $x$ is also illustrated in Fig.~\ref{fig:R}.

Many of the analyses listed in Table~\ref{tab:table} imposed a flavor independence of the valence and light sea quarks at the parametrization scale $Q^2=Q^2_0$,
\begin{equation}
R^A_{u_{\rm V}}(x,Q^2_0)    = R^A_{d_{\rm V}}(x,Q^2_0), \quad \quad
R^A_{\overline{u}}(x,Q^2_0) = R^A_{\overline{d}}(x,Q^2_0) = R^A_{\overline{s}}(x,Q^2_0).
\end{equation}
This has been (and still is) perfectly consistent with the data. However, it is known that the above symmetry is quickly destroyed at higher $Q^2$ by the DGLAP evolution such that there is no real reason to assume it in the first place. To this end, the nCTEQ15 analysis allowed some flavor variation for the valence quarks, but the new EPPS16 fit went one step ahead and let also the sea quarks to be flavor dependent (small and intermediate $x$). By doing this, the imposed bias is much reduced in comparison, say, to EPS09 or DSSZ.

\begin{wrapfigure}{r}{0.600\textwidth}
\hspace{-0.6cm}
\includegraphics[width=1.10\linewidth]{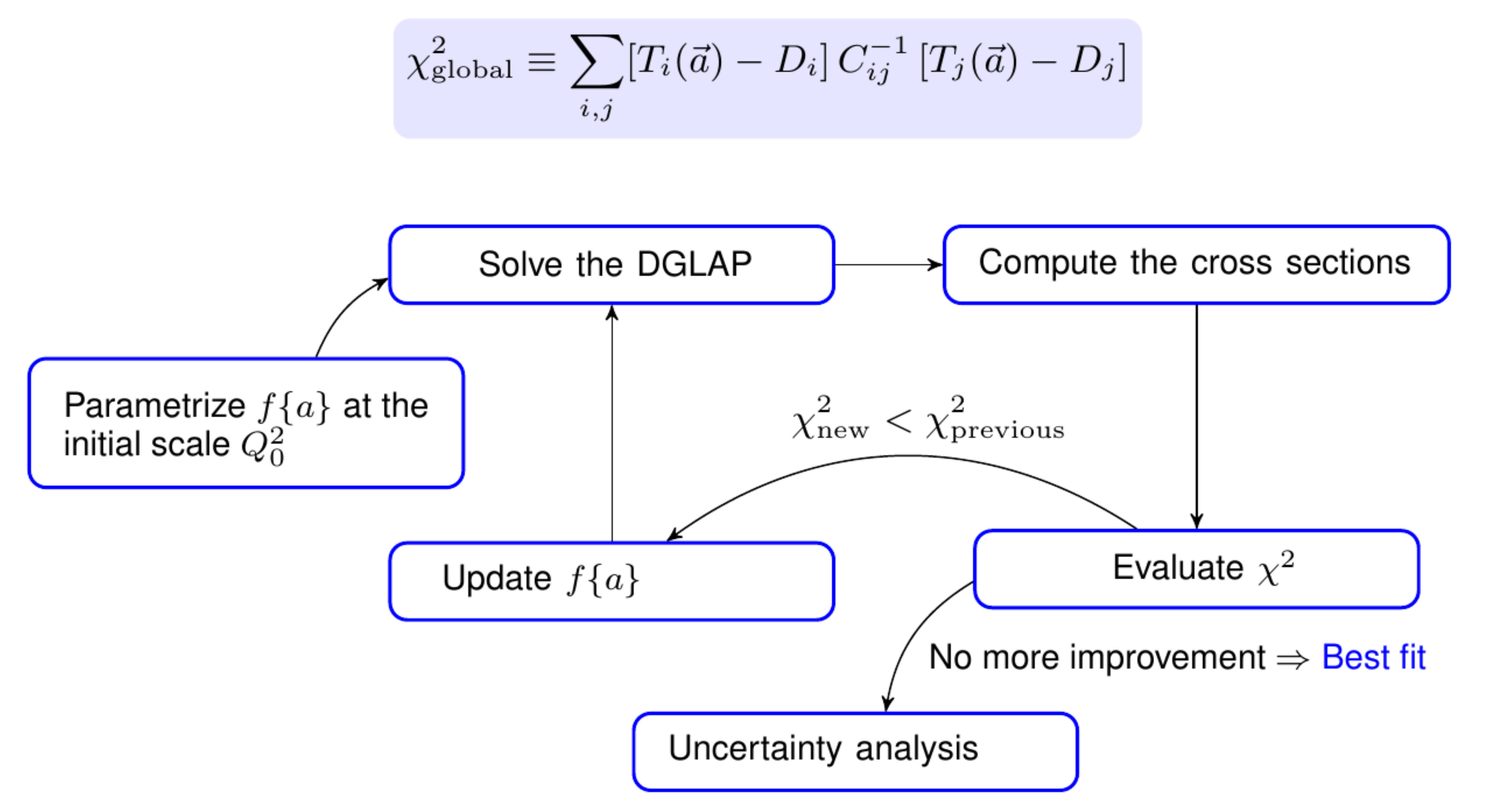}
\caption{The steps of a typical global analysis. In the definition of $\chi^2_{\rm global}$, $T_i(\vec a)$ denotes the theory value corresponding to the $i$th data point, $D_i$ is the corresponding measured value, and $C_{ij}^{-1}$ is the inverted covariance matrix.}
\label{fig:steps}
\end{wrapfigure}

All the current analyses adopt the following fit procedure which is also depicted in Fig.~\ref{fig:steps} here: The PDFs are first parametrized with some variables $\vec a$ at a low scale $Q^2=Q^2_0$, and one solves the DGLAP evolution to higher $Q^2$ and computes the cross sections corresponding to the input data. One can then evaluate the global $\chi^2_{\rm global}$ figure-of-merit function which measures the agreement between the data and the theory. The $\chi^2_{\rm global}$ function is minimized iteratively and the found minimum is defined as the central fit. 

The usual way to estimate the PDF uncertainties is the Hessian method \cite{Pumplin:2001ct}. This technique is based on expanding $\chi^2_{\rm global}$ around the minimum $\chi^2_0$ to second order in the fit-parameter variations $\delta a_i$ as
\begin{equation}
\chi^2_{\rm global} \approx \chi^2_0 + \sum_{i,j} \delta a_i
        H_{ij}
        \delta a_j = \chi^2_0 + \sum_{i} z_i^2 \, .
\end{equation}
By diagonalizing the the second-derivative matrix $H_{ij}$ one gets largely rid of the inter-parameter correlations and in the ``diagonalized'' $z$ coordinates it is justified to use the standard law of error propagation to compute the uncertainty $\delta X$ for any quantity $X$ that depends on the PDFs,
\begin{equation}
(\delta X)^2 = \sum_i \left( \frac{\partial X}{\partial z_i} \times \delta z_i \right)^2, \ \delta z_i= \frac{\delta z_i^++\delta z_i^-}{2} \, .
\end{equation}
The usefulness of the Hessian procedure is in the definition of uncertainty sets $S_i^\pm$, in $z$-space coordinates,
\begin{equation}
S_1^\pm \equiv \pm \delta z_1^\pm \left(1,0,\ldots,0 \right), \ S_2^\pm \equiv \pm \delta z_2^\pm \left(0,1,\ldots,0 \right), \, \ldots \, , S_N^\pm \equiv \pm \delta z_N^\pm \left(0,0,\ldots,1 \right) \, ,
\end{equation}
as, by doing this, the propagation of the fit uncertainties to any observable boils down to evaluating it with the error sets (for an asymmetric prescription, see e.g. Refs.~\cite{Eskola:2009uj,Eskola:2016oht}),
\begin{equation}
(\delta X)^2 = \frac{1}{4} \sum_i \left[ X(S_i^+) - X(S_i^-)\right]^2 \, .
\end{equation}

Perhaps the largest conceptual difficulty in the Hessian method is how to determine the parameter variations $\delta z_i^\pm$ that go in the definition of error sets. All the current fits define them to correspond to a specific global tolerance $\Delta \chi^2$. Ideally, this is just unity, but essentially for the parametrization bias, the global fits need to use much larger numbers, see Table~\ref{tab:table}. In EPS09, EPPS16 and nCTEQ15 analyses these figures are based on the so-called dynamical tolerance determination \cite{Martin:2009iq} requiring that all the data are, on average, reproduced within a 90\% confidence level. That is, the process is data driven.

\vspace{-0.3cm}
\section{The use of experimental data is not unambiguous}
\label{Theuseofexperimentaldataisnotunambiguous}

\begin{wrapfigure}{r}{0.450\textwidth}
\hspace{-0.2cm}
\centerline{
\includegraphics[width=1.05\linewidth]{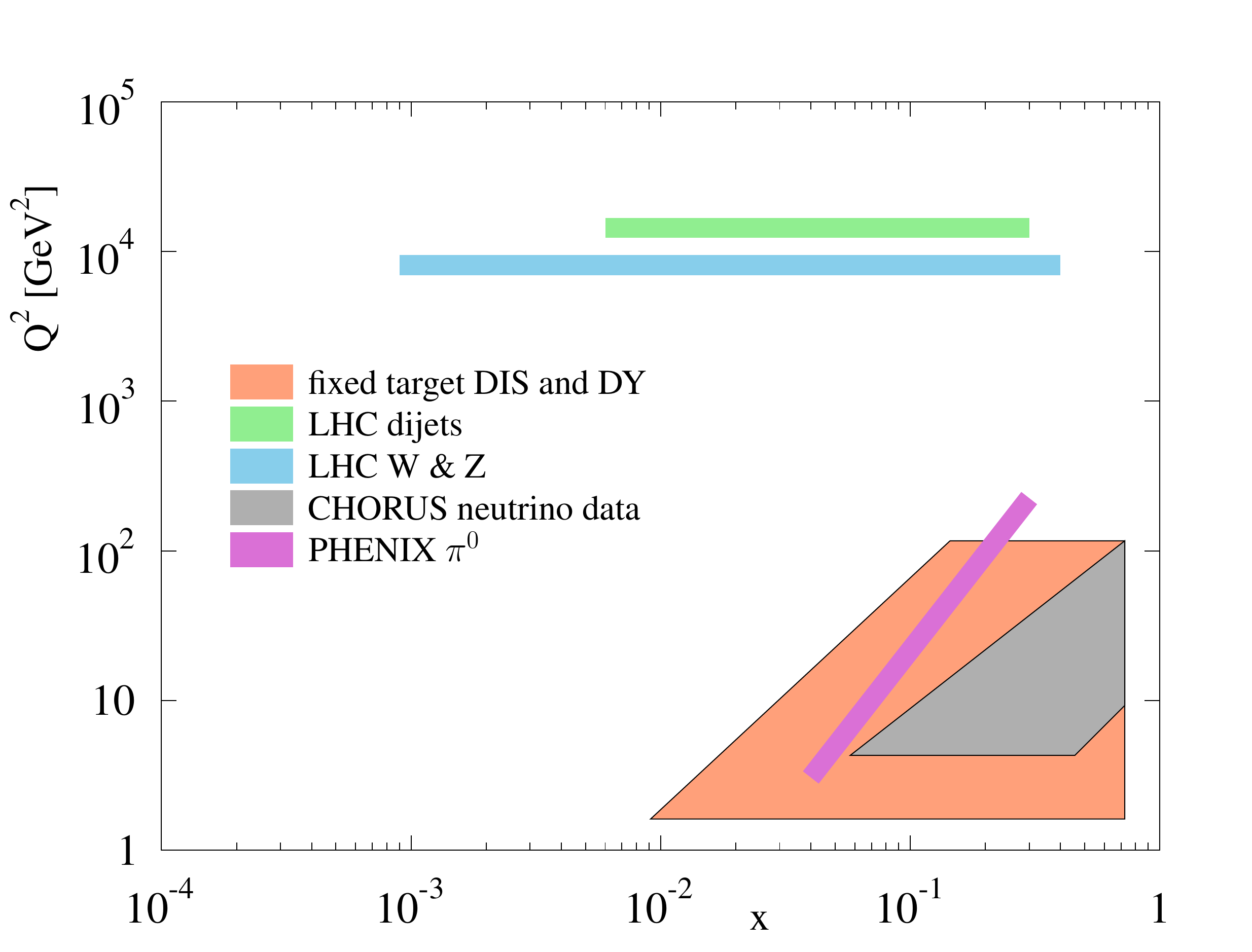} \\
}
\label{fig:xQ}
\caption{The kinematical coverage of the data used in nuclear-PDF fits. Figure from Ref.~\cite{Eskola:2016oht}.}
\end{wrapfigure}

\vspace{-0.2cm}
Fig.~\ref{fig:xQ} shows the kinematic coverage of the data presently used in global fits. All the old fixed-target and RHIC pion data are in the lower right-hand corner. The new LHC p--Pb data probe the nuclear PDFs in a completely different kinematical domain thereby significantly increasing the kinematical reach of constraints. How should these new LHC data be used to find information on the nuclear modifications? One possibility is to use directly the measured absolute cross sections. However, the problem then is that the absolute spectra are very sensitive to the used proton PDFs which enter also from the nuclear-PDF side, as discussed earlier. As a consequence, the interpretation of the data is ambiguous. This approach was nevertheless used in a recent PDF-reweighting work by the nCTEQ collaboration \cite{Kusina:2016fxy}. The second option is to use the cross sections normalized to the rapidity-integrated cross section, in which case part of the free-proton uncertainty indeed cancels out. However, the cancellation is not particularly complete as can be appreciated e.g. from a recent study in Ref.~\cite{Armesto:2015lrg} (see Table 3 there). The third and best option is to use the forward-to-backward ratios in which as much as possible of the free-proton PDF dependence cancels. The cancellation is still not complete, however. One also cancels some experimental uncertainties, the luminosity above all, but also loses some information. For example, the forward-to-backward ratio is always close to unity near the midrapidity, by construction. Clearly, the Pb--Pb data cannot be used in this way. The forward-to-backward ratio usually differs from unity for several reasons: nuclear modifications in PDFs, phase-space effects, and isospin effects.

There are several measurements on neutrino DIS (NuTeV, CCFR, CHORUS, CDHSW, Minerva) and, apart from the very recent low-energy measurements of Minerva, the data are available only as absolute cross-sections. Thus, one encounters a similar difficulty as in the case of LHC data: the data are sensitive to both the free-proton baseline and the nuclear modifications. In the works of nCTEQ \cite{Kovarik:2010uv} and DSSZ \cite{deFlorian:2011fp} these absolute cross-sections are nevertheless used and nCTEQ has even reported a significant tension with the other measurements \cite{Kovarik:2010uv}. To reduce theoretical bias and also the experimental uncertainties, it was proposed \cite{Paukkunen:2013grz} that cross-sections normalized to the integrated ones should be used instead. Indeed, this approach then revealed the usual pattern of antishadowing and EMC effect in the neutrino data from various independent collaborations. It is in this way that the neutrino DIS data are now incorporated in the EPPS16 analysis, accounting also for the correlated systematic uncertainties.

There are also ambiguities related to the old fixed target DIS data. The original measurements were corrected for the isospin effects by the experiments, as if the measurements had involved only isoscalar nuclei. These corrected data have been used in majority of the available fits. A better alternative is to use the original, non-isospin-corrected versions of the measurements which is the approach adopted in the EPPS16 fit. This option removes the bias caused by the assumptions made on the isospin corrections.

\begin{figure}
\centering
\includegraphics[width=0.25\linewidth]{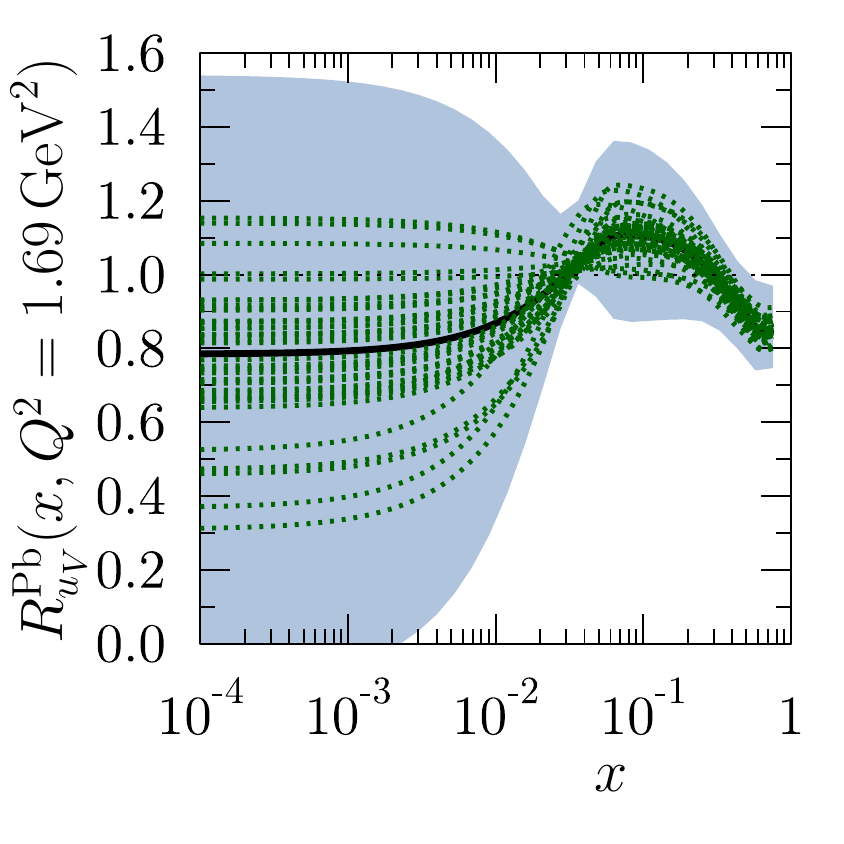}
\includegraphics[width=0.25\linewidth]{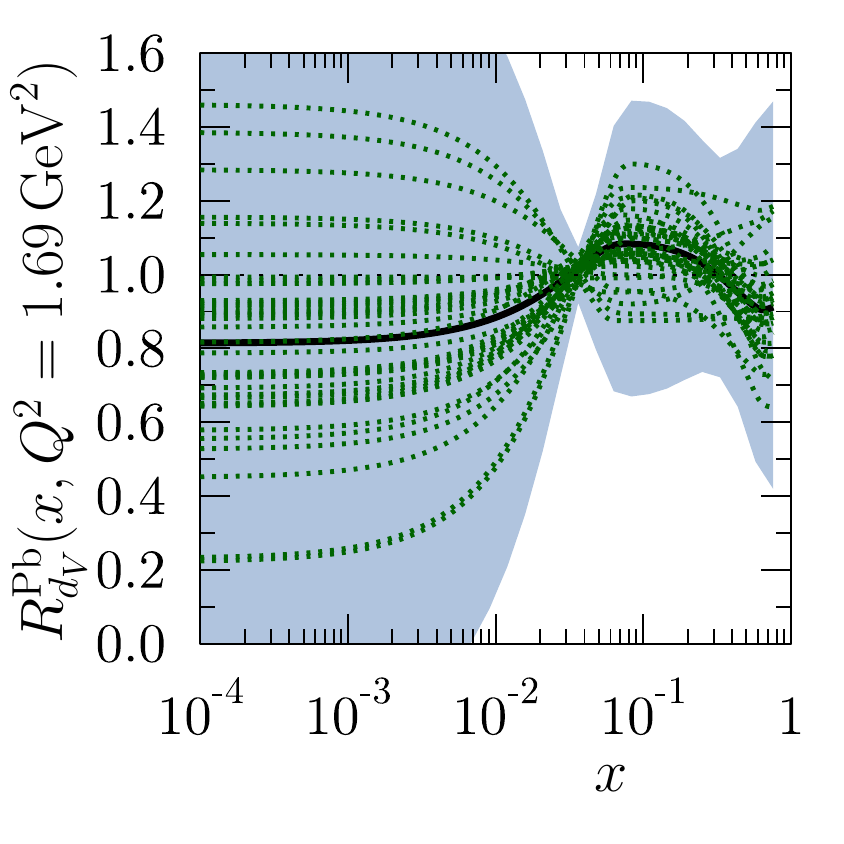}
\includegraphics[width=0.25\linewidth]{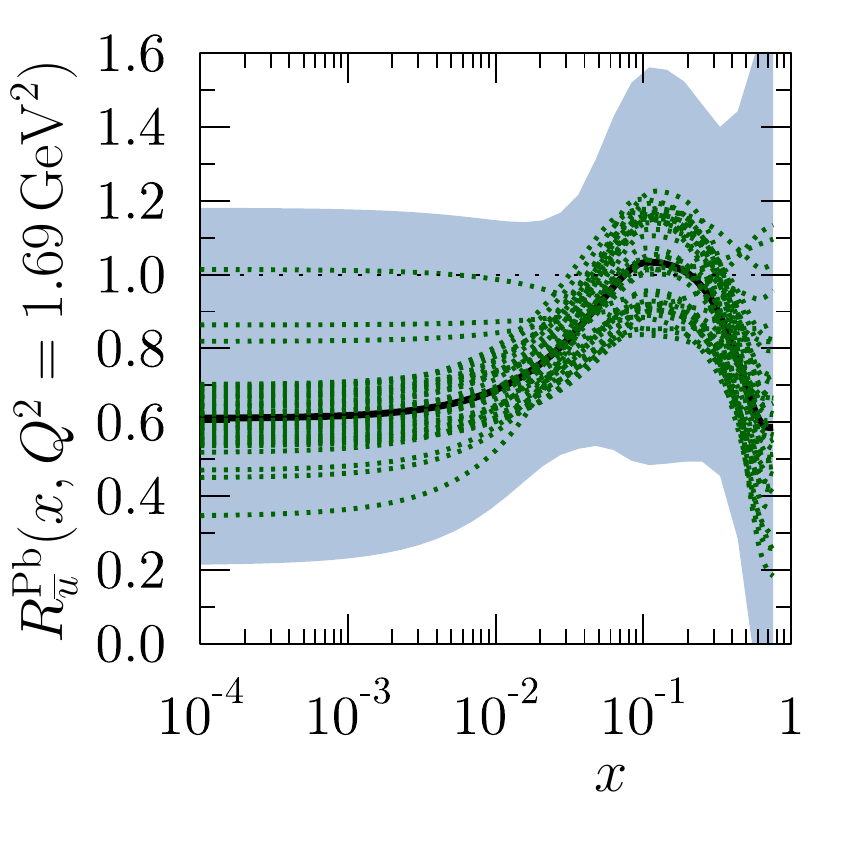} \\
\includegraphics[width=0.25\linewidth]{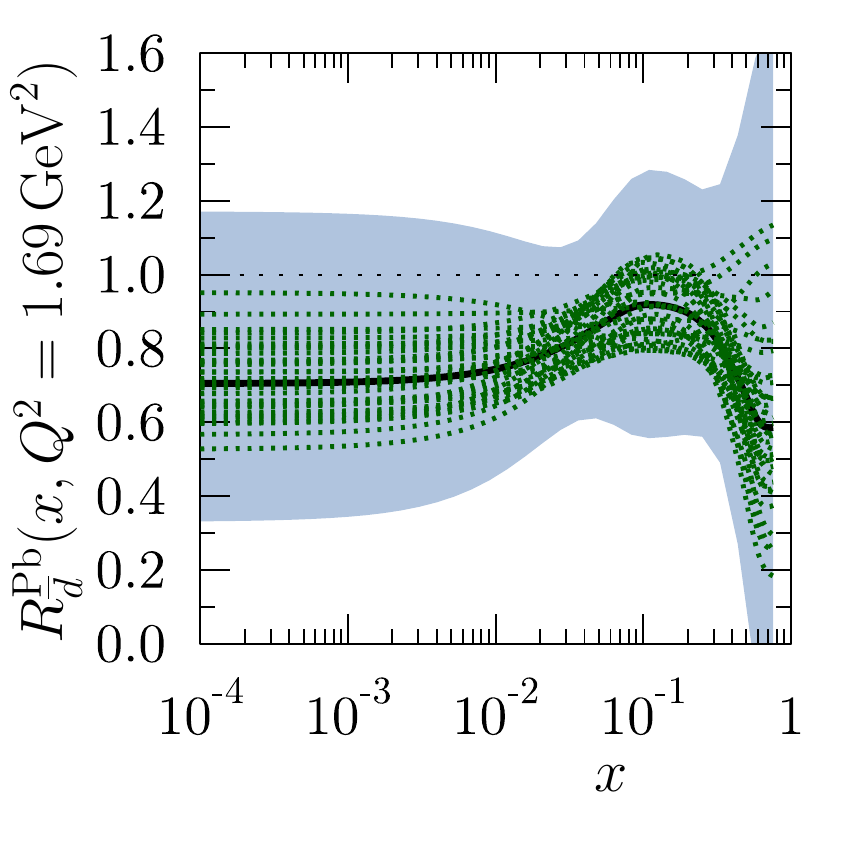}
\includegraphics[width=0.25\linewidth]{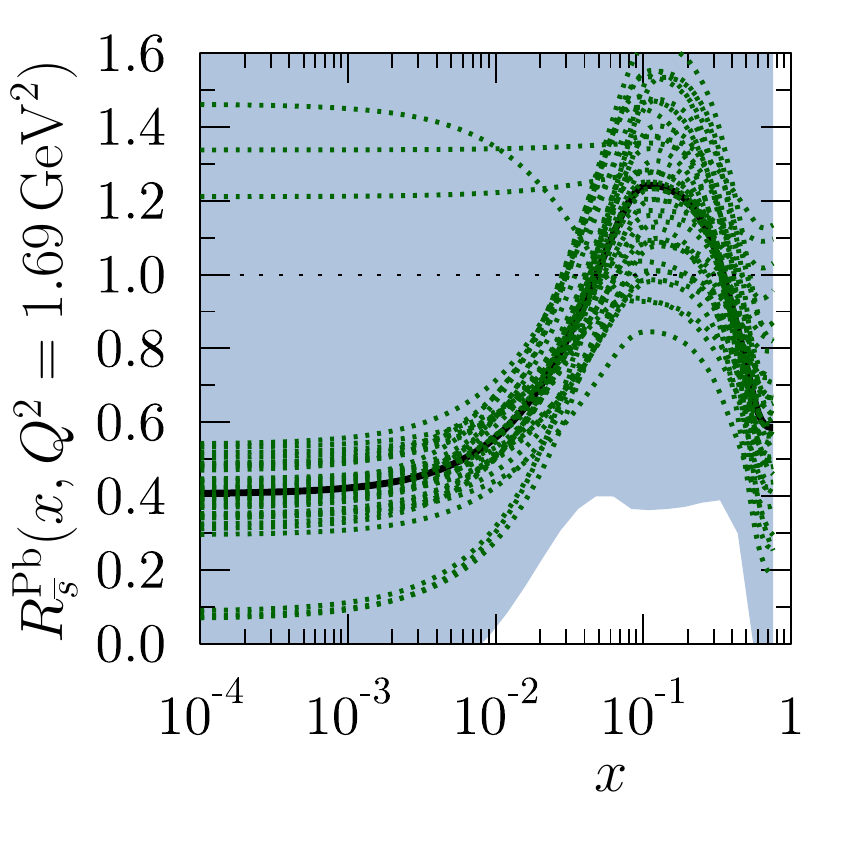}
\includegraphics[width=0.25\linewidth]{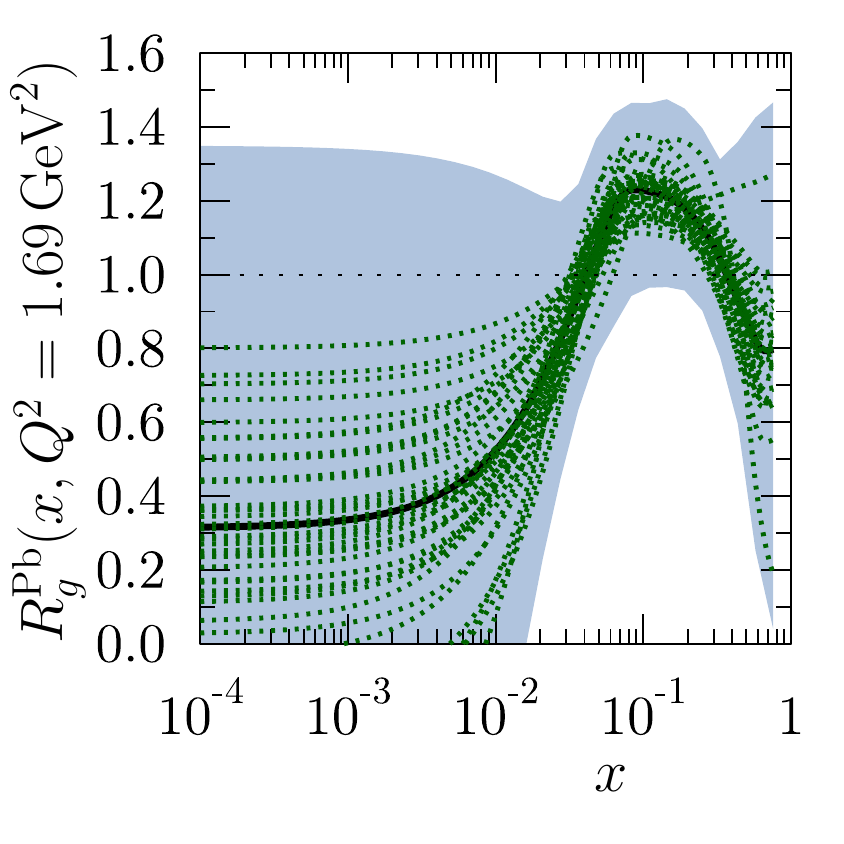}
\caption{The EPPS16 nuclear modifications at the parametrization scale $Q^2=1.69 \, {\rm GeV}^2$. The central results are shown as black curves and the individual error sets $S_i^\pm$ are green dotted curves. The total 90\% uncertainty is shown as blue bands. Figure from Ref.~\cite{Eskola:2016oht}.}
\label{fig:EPPS16lowQ}
\end{figure} 

\vspace{-0.3cm}
\section{Comparison of the current global fits}
\label{Comparisonofthecurrentglobalfits}

\begin{figure}[htb!]
\centering
\includegraphics[width=0.95\linewidth]{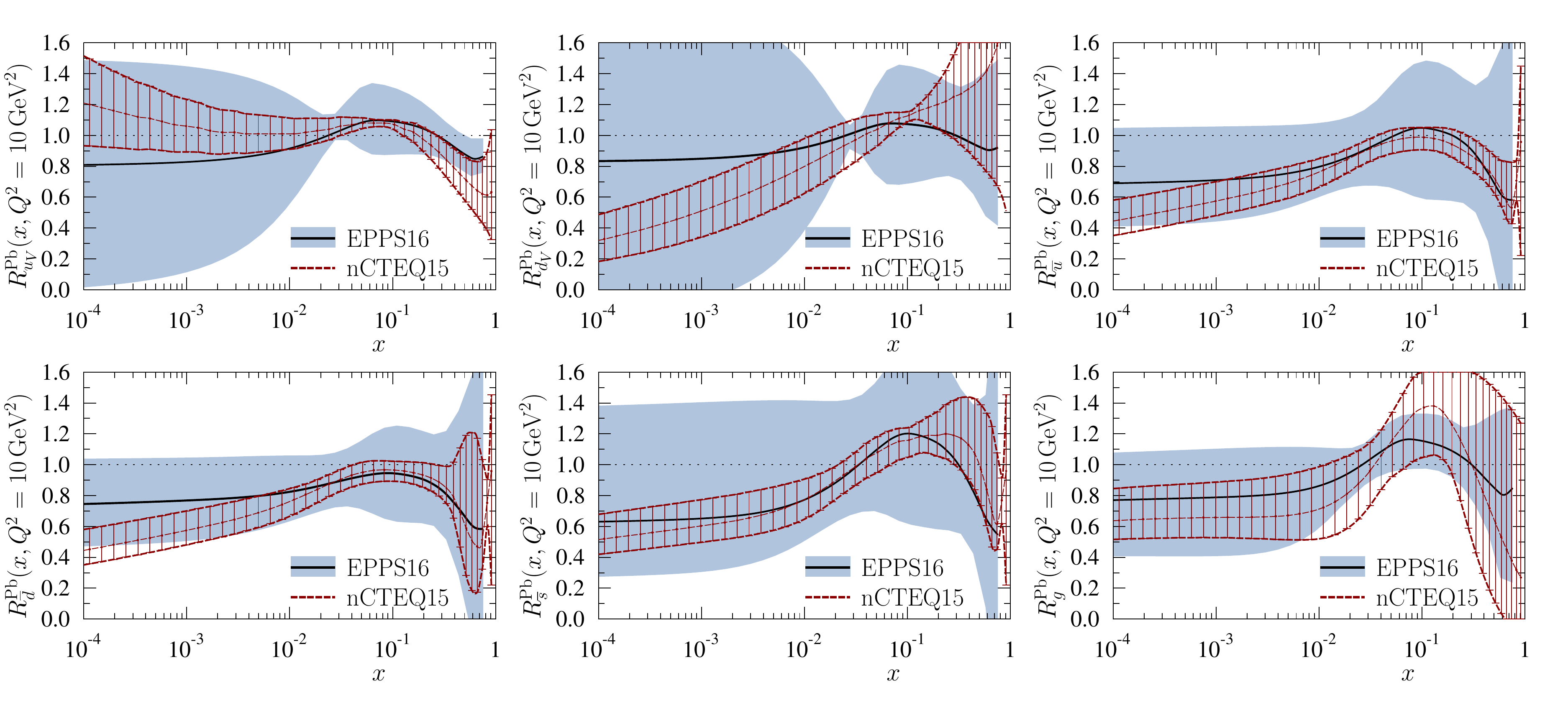}
\caption{The EPPS16 nuclear modifications at $Q^2=10 \, {\rm GeV}^2$ compared to the results of nCTEQ15 analysis. Figure from Ref.~\cite{Eskola:2016oht}.}
\label{fig:EPPS16nCTEQ}
\end{figure}

\vspace{-0.2cm}
Let us then compare the current global fits starting with EPPS16 at the charm-mass threshold, shown in Fig.~\ref{fig:EPPS16lowQ}. For their smallness at small $x$, the valence-quark distributions can only be constrained at relatively large $x$, and there both flavors show mutually a rather similar behavior as far as the best fit is concerned (antishadowing + EMC effect). The sea quarks are better under control at small $x$ apart from the strange quarks for which the uncertainty is enormous. The gluons are relatively well constrained at large $x$, but at small $x$ the uncertainties are large. However, the uncertainties quickly diminish when moving to higher $Q^2$. The situation near the J/$\psi$ mass scale is shown in Fig.~\ref{fig:EPPS16nCTEQ} where also the results from nCTEQ15 analysis are overlaid. The nCTEQ15 errors are usually clearly smaller than those of EPPS16, which follows essentially from the more restrictive assumptions made in nCTEQ15. The high-$x$ gluon uncertainties are, however, larger in nCTEQ15 since no LHC data are included and since the cuts for DIS data are more restrictive. The mutual behavior of the nCTEQ15 up and down valence is also different from EPPS16. This is most likely related to the use of isospin-corrected DIS data and for not including any neutrino DIS data in nCTEQ15.

\begin{figure}[htb!]
\centering
\includegraphics[width=0.95\linewidth]{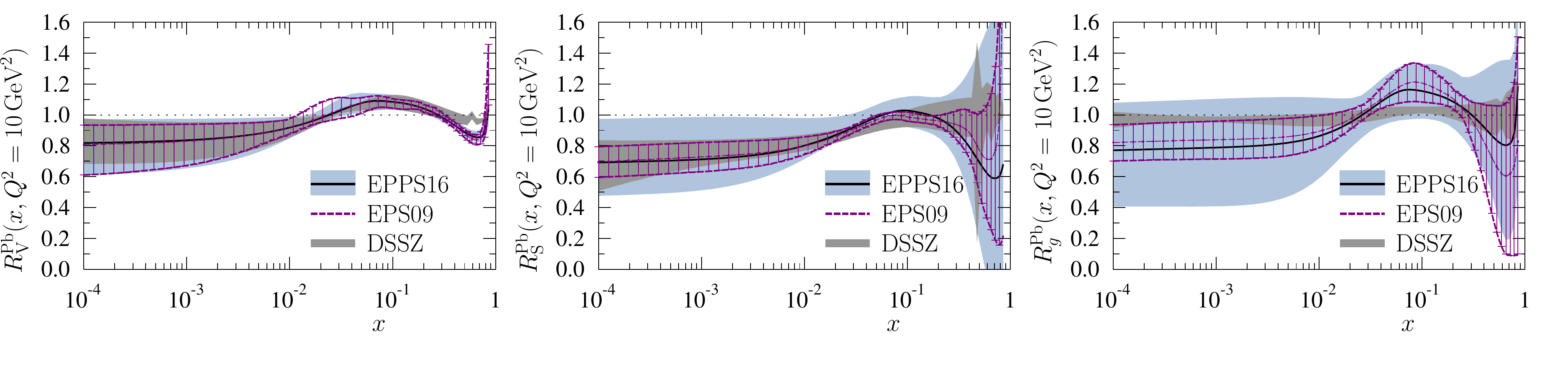}
\caption{The EPPS16 nuclear modifications at $Q^2=10 \, {\rm GeV}^2$ compared to the EPS09 and DSSZ analyses. Figure from Ref.~\cite{Eskola:2016oht}.}
\label{fig:EPPS16EPS09}
\end{figure}

A comparison between EPPS16, EPS09 and DSSZ is shown in Fig.~\ref{fig:EPPS16EPS09}. Both the EPS09 and DSSZ fits did impose the flavor-independence of the quark nuclear effects at the parametrization scale (discussed in Section~\ref{Analysisprocedures}) and I therefore compare only flavor averages
\begin{equation}
R_V^{\rm Pb}  \equiv  \frac{u^{\rm p/Pb}_{\rm V}+d^{\rm p/Pb}_{\rm V}}{u^{\rm p}_{\rm V}+d^{\rm p}_{\rm V}}, \quad \quad
R_S^{\rm Pb} \equiv \frac{\overline{u}^{\rm p/Pb}+\overline{d}^{\rm p/Pb}+\overline{s}^{\rm p/Pb}}{\overline{u}^{\rm p}+\overline{d}^{\rm p}+\overline{s}^{\rm p}}.
\end{equation}
All the three are consistent with each other, modulo the valence quarks of DSSZ at very high $x$ where there is probably an issue with the isoscalar correction. The EPPS16 uncertainty bands are broader for the larger amount of freedom in the fit functions, even though the analysis contains more data than the other two.

\begin{figure}[htb!]
\centering
\includegraphics[width=0.350\linewidth]{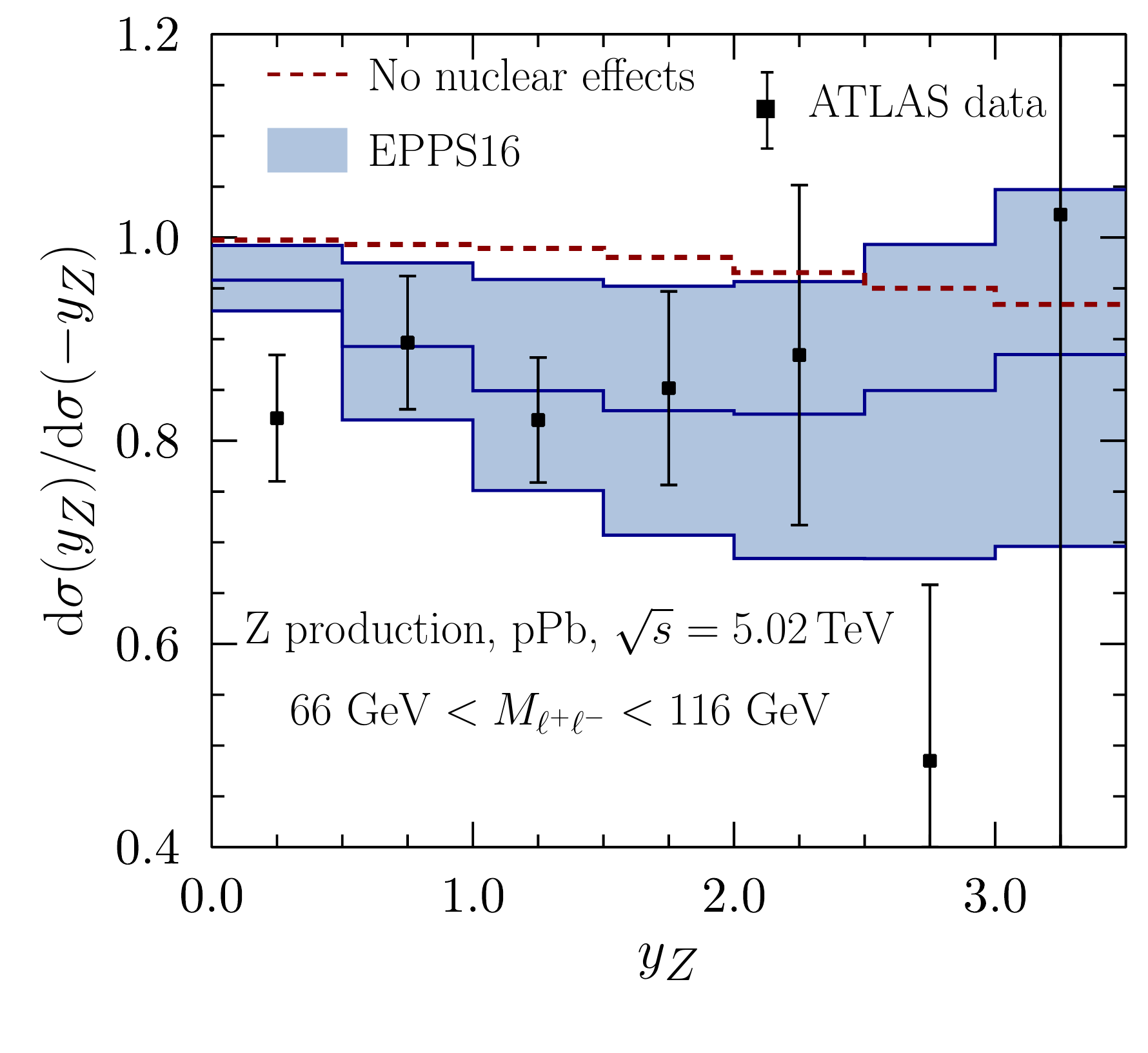}
\includegraphics[width=0.350\linewidth]{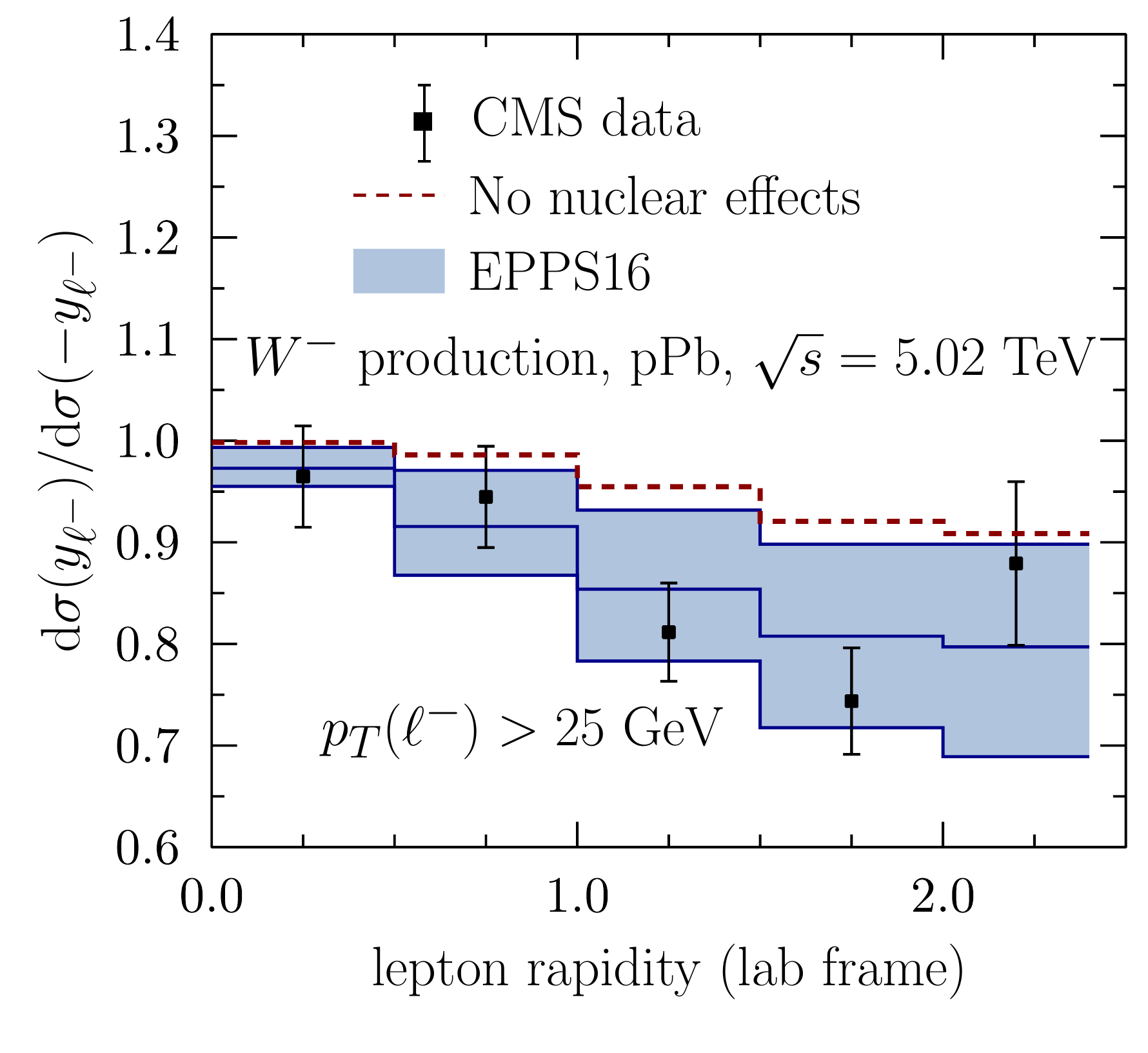}
\includegraphics[width=0.350\linewidth]{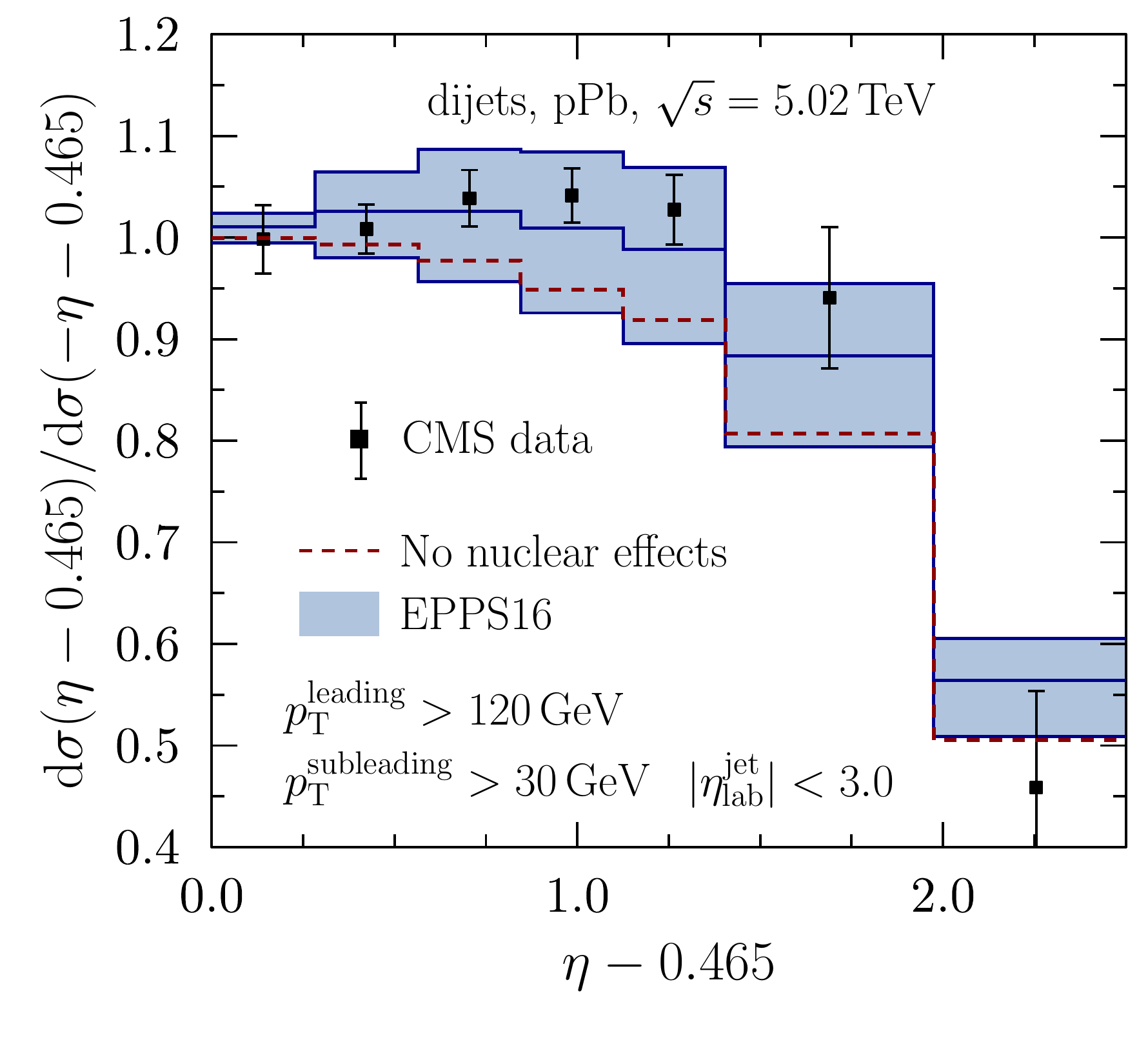}
\includegraphics[width=0.350\linewidth]{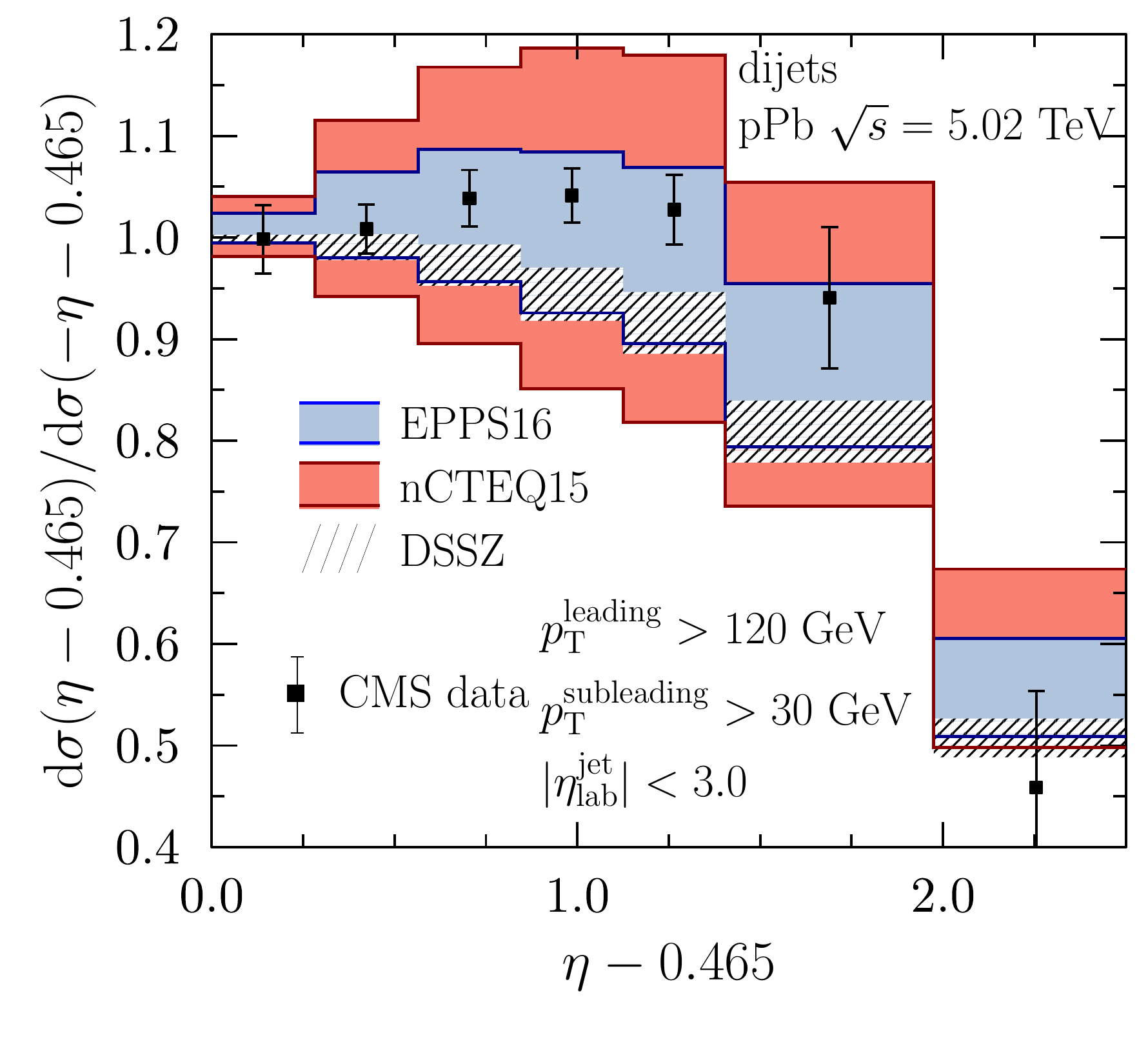}
\caption{Upper panels: The forward-to-backward ratios for Z (left) and W$^-$ (right) production in LHC p--Pb collisions compared to EPPS16. The data are from the ATLAS \cite{Aad:2015gta} and CMS \cite{Khachatryan:2015hha} collaborations. Lower panels: The CMS dijet data \cite{Chatrchyan:2014hqa} compared with the EPPS16 fit (left) and also with the results obtained using the nCTEQ15 and DSSZ parametrizations (right). Figure from Ref.~\cite{Eskola:2016oht}.}
\label{fig:WZ}
\end{figure} 

\vspace{-0.3cm}
\section{Effects of nuclear PDFs in LHC p--Pb observables}
\label{EffectsofnuclearPDFsinLHCpPbobservables}

\vspace{-0.2cm}
What kind of evidence of nuclear-PDF effects do the LHC data entail? As an example, Fig.~\ref{fig:WZ} presents (upper panels) forward-to-backward ratios for Z and W$^-$ as included in the EPPS16 fit. Since there is more net shadowing at forward direction than in the backward direction, the forward-to-backward ratios acquire a suppression in comparison to the calculation with only proton PDFs (red dashed lines). The data tend to support this behavior.

Unlike the Z and W production, the CMS dijet measurement \cite{Chatrchyan:2014hqa} show evidence for an enhancement in the forward-to-backward ratio when compared to the calculation with only proton PDFs (red dashed lines). This is shown in Fig.~\ref{fig:WZ} (lower panels). In EPPS16 this is explained, in essence, by the antishadowed gluons ($\eta>0$) getting divided by EMC suppressed gluons ($\eta<0$). The data deviates quite much from unity at large dijet pseudorapidity $\eta$ for the non-symmetric acceptance in the nucleon-nucleon center-of-mass frame. In Fig.~\ref{fig:WZ}, I also compare the dijets with other sets of nuclear PDFs. The nCTEQ15 uncertainties for high-$x$ gluons are larger (see Fig.~\ref{fig:EPPS16nCTEQ}) and correspondingly the error band for the dijets is wider as well. The DSSZ parametrization does not reproduce these data. 

\vspace{-0.3cm}
\section{The way forward}
\label{Thewayforward}

\vspace{-0.2cm}
In the near future, I would expect much new data on W, Z, jet, photon, top-quark, and heavy-flavor measurements from p--Pb and Pb--Pb collisions at the LHC. Direct measurements of the nuclear modification $R_{\rm pPb}$ are now possible for the new p-p baseline run at $\sqrt{s}=5\,{\rm TeV}$. More or less so also with the new $\sqrt{s}=8.16\,{\rm TeV}$ p--Pb data, though the p-p baseline is not exactly at the same $\sqrt{s}$. While the $R_{\rm pPb}$s would nicely reveal the nuclear effects (if the isospin effects are not large) it is important to account for the correlated systematic uncertainties between the p-p and p--Pb measurements --- the detector is the same, so the two must be correlated. Also, it would be advantageous to perform the measurements in the nucleon-nucleon center-of-mass frame with symmetric rapidity cuts (see e.g. Ref.~\cite{Eskola:2013aya}, Fig.~6, for a case study on dijets). 

The low- and intermediate-mass Drell-Yan production has not yet been measured at the LHC p--Pb collisions, but the prospects are that this process could have some discriminating power for the quark distributions \cite{Arleo:2015qiv}. This should be within the possibilities of e.g. the LHCb experiment at forward direction \cite{LHCb-PUB-2016-011}. There is also new low-mass Drell-Yan measurement soon coming from the Fermilab SeaQuest experiment \cite{Dannowitz:2016qkz} which will be interesting. The J/$\psi$ production is theoretically less robust and some, say, nuclear absorption or energy loss may be involved \cite{Arleo:2015qiv}. In a recent study \cite{Lansberg:2016deg} the idea was not to take a stand on the actual J/$\psi$ formation process, but instead assume that it is the gluon-gluon partonic channel that dominates, and fit the coefficient functions to p-p data. The predictions for p--Pb are then obtained by simply switching the PDFs. Interestingly, the authors obtain a consistent description of the current p--Pb data with only effects from nuclear PDFs. The potential of open heavy flavor in constraining the PDFs has been recently demonstrated in p-p collisions \cite{Zenaiev:2015rfa,Gauld:2016kpd}. Indeed, it has been shown that a huge reduction in the gluon uncertainty can be obtained by including LHCb D and B meson data. While the theoretical description is not unique, it is claimed that in different ratios much of the uncertainties tend to cancel out. Thus, the prospects for introducing D and B meson data from p--Pb runs \cite{LHCb:2016huj} also in nuclear-PDF studies look promising. It has also been argued that the exclusive vector meson production in ultraperipheral Pb--Pb collisions should serve as a strong constraint for the nuclear gluons. The prediction is that the cross sections scales as the gluon distribution squared \cite{Ryskin:1992ui}, and the available data appear to indeed favor a shadowing similar to that in EPS09, see e.g. Ref.~\cite{Thomas:2016oms}. However, there is a certain question mark on how exactly does the gluon distribution probed by this \emph{exclusive} process correspond to the usual, \emph{inclusive} (NLO and beyond) PDFs.

\vspace{-0.3cm}
\section{Summary}
\label{Summary}

\vspace{-0.2cm}
I have overviewed the situation of global fits of nuclear PDFs as it stands. The most important recent developments are the inclusion of LHC p--Pb data on W, Z, and jet production (realized currently only in EPPS16), the less biased incorporation of neutrino DIS data which has an important impact on the valence distributions (realized currently only in EPPS16), and freeing the flavor dependence of the nuclear modifications which leads to much less bias, but also increases the uncertainties flavor by flavor. It can be expected that new LHC measurements will be available very soon in the future. To this end, I would like to take the opportunity to stress the importance of making the correlated experimental bin-by-bin systematics available also in nuclear collisions --- in p-p case this has been the usual practice already for several years. This would increase the impact of the data especially now that the statistics are higher in the new $\sqrt{s}=$8.16\,TeV p--Pb data sample and the role of systematic errors thereby more pronounced. Also, to reduce theoretical uncertainties, it would be advantageous to perform the p--Pb measurements with a symmetric rapidity acceptance in the nucleon-nucleon center-of-mass frame.

\vspace{-0.4cm}
\section*{Acknowledgments}

\vspace{-0.2cm}
I acknowledge the funding from Academy of Finland, Project 297058 ; the European Research Council grant HotLHC ERC-2011-StG-279579 ; Ministerio de Ciencia e Innovaci\'on of Spain and FEDER, project FPA2014-58293-C2-1-P; Xunta de Galicia (Conselleria de Educacion) - H.P. is part of the Strategic Unit AGRUP2015/11.

\vspace{-0.4cm}
\bibliographystyle{elsarticle-num}
\bibliography{HP}

\end{document}